\documentclass[preprint,12pt]{elsarticle}
\usepackage{amsmath}
\usepackage{txfonts}
\usepackage{amsmath,amsfonts,amssymb,graphicx,mathrsfs,epsfig,subfigure}
\usepackage[dvipdfm,colorlinks,linkcolor=blue,citecolor=blue]{hyperref}
\usepackage{multirow}
\usepackage{mathrsfs}
\usepackage{booktabs}
\usepackage{fancyhdr,times}
\usepackage{graphicx}
\newtheorem{Lemma}{\hskip\parindent\bf{Lemma}}[section]

\newtheorem{Remark}{\hskip\parindent\bf{Remark}}[section]
\newtheorem{Theorem}{\hskip\parindent\bf{Theorem}}[section]
\biboptions{sort&compress}

\begin{document}

\begin{frontmatter}

\title{\Large\bf Thresholds and bistability in HIV infection models with oxidative stress \tnoteref{t1}}
 \tnotetext[t1]{This work is supported by NSFC (Nos. 11671346 and U1604180), NSF of Henan Province (No. 162300410031) and Nan Hu
Scholar Development Program of XYNU. }
\author[els]{Shaoli Wang \corref{cor1}}
\ead{wslheda@163.com }
\author[wlu]{Fei Xu}
\ead{fxu.feixu@gmail.com}
\author[Song]{ Xinyu Song  }
\ead{xysong88@163.com}

 \cortext[cor1]{Corresponding author.}

\address[els]{School of Mathematics and Statistics, Henan University, Kaifeng 475001, Henan, PR China }
\address[wlu]{Department of Mathematics, Wilfrid Laurier University, Waterloo, Ontario, Canada \ N2L 3C5}
\address[Song]{College of Mathematics and Information Science, Xinyang Normal University, Xinyang, 464000, Henan, PR China}

\begin{abstract}
% Text of abstract

 Oxidative stress, a reaction caused by the imbalance between the
reactive oxygen species of human organism and its ability to
 detoxify  reactive intermediates and to repair the resulting damage plays an important role in HIV-infections. On one hand,
 HIV infection is responsible for the chronic oxidative stress of the patients. On the other hand, the oxidative stress
  contributions to the  HIV disease pathogenesis.  In this paper, we integrate oxidative stress into an HIV infection model to investigate
   its effects on the virus dynamics. Through mathematical analysis, we obtain the basic reproduction number $R_{0}$ of the model
   which describes the persistence of viruses. In particular, we show that for $R_{0}>1$, the model has a bistable interval with virus
   rebound threshold and elite control threshold. Numerical simulations and bifurcation analysis are presented to illustrate the viral
   dynamics under oxidative stress. Our investigation reveals the interplay between viruses and the reaction of human organism including
   immune response and oxidative stress, and their effects on the health of human being.

\end{abstract}


\end{frontmatter}

\section{Introduction}

Combination antiretroviral therapy gives patients long-term
suppression of HIV with undetectable viral levels. Recent
investigations show that for patients with undetectable viral
levels, there exists a reservoir in which viruses remain alive in a
long-lived latent state. With the termination of receiving
combination antiretroviral therapy, after an average time period of
18 days, plasma viremia rebounds to detectable level. In the
literature, medical cases of HIV rebound were reported.
%
%The
%`Mississippi baby' got HIV from her HIV-positive birth mother and
%received liquid, triple-drug antiretroviral treatment at the age of
%30 hours \cite{NIH}. A medical examination at the age of about 23
%months indicated an undetectable viral level. Though doctors thought
%the `Mississippi baby'  has been cured of HIV,
%
Medical examinations for the `Mississippi baby'
at the age of nearly 4 years old displayed a rebounded HIV level in
blood (16,750 copies/ml) \cite{NIH}.
The `Mississippi baby' case
implies that there exists a time delay for the viral rebound
\citep{Persaud,NIH}. Researches had been carried to explain such
phenomena \citep{Persaud,Conway}.

 Conway¡¡and Perelson \cite{Conway}
 developed a mathematical system to model to study  the
dynamics of HIV infection. Their investigation captures the
interplay between viral dynamics and the response of the host and
brings insight into the evolution of the disease infection process.
By evaluating the designed model, the authors predicted that the
strength of immune response and the initial size of the latent
reservoir could affect the dynamics. Their results on post-treatment
control provide guidance for future studies. Investigations implies
that after HIV infection, patients who receive antiretroviral
therapy early have a higher chance of getting post-treatment
control, a situation where the amount of plasma virus remains
undetectable after the termination of the medical treatment.
However, clinically, only a small proportion of such patients that
receive early treatment attain post-treatment control. Further
investigations are to be carried out to reveal the mechanism behind
the post-treatment control.

As a highly reactive oxygen species (ROS), oxidants are continuously
produced during normal biochemical reactions in the human body. Most
cells are able to detoxify physiologic levels of ROS in the human
body using antioxidants such as enzymes. The processes of producing
oxidants and detoxication may reach an equilibrium state. When the
balance between the two processes disturbed, a condidtion called
oxidative stress occurs \cite{TangAM}.  Antioxidants play an
important role in regulating the reactions that release free
radicals. Cells need to reach certain level of antioxidant defenses
to counteract the detrimental effects caused by an excessive
production of ROS to protect the immune system \cite{Puertollano}.
Investigations suggest the existence of interactions between
HIV-infection and oxidative stress. On one hand, the HIV-infection
process contributes to the disturbance of the balance between the
generation of free radicals and antioxidant defenses. On the other
hand, oxidative stress is beneficial to HIV disease pathogenesis by
promoting the replication of viruses, decreasing the proliferation
of immune cells, and increasing the sensitivity to drug toxicities
etc \cite{Pace}.

 After HIV infection, the initial reaction of
the host is rapid and nonspecific by activating natural killer
cells, macrophage cells, etc. The host develops delayed and specific
reactions by activating CTLs and antibody cells. During most viral
infections, CTLs attack infected cells and antibody cells attack
viruses. These attacks against the viruses act as an antiviral
defense for the host. Dynamics of viral infection with CTL response
have been investigated in the literature. \cite {Nowak1} introduced
a mathematical system to model the  interplay  between activated
CD4$^{+}$ T cells, infected CD4$^{+}$ T cells, viruses and immune
cells. \cite{Bartholdy} and \cite{Wodarz} performed investigation on
the HIV infection and concluded that the turnover of free virus is
much faster than that of infected cells. Based on these results,
they proposed the quasi-steady state assumption, i.e., the load of
free virus is proportional to the amount of infected cells. Thus we
can estimate the viral load by evaluating the number of infected
cells.

Investigations demonstrated that HIV mutates into new forms that
escape from specific immune responses or immune exhaustion during
virus evolution
\citep{Nowak2,Brrow,Goulder,Price,Phillips,Kaufmann,Khaitan,Johnson}.
HIV infection may modulate dendritic cells, which are responsible
for the viral evasion from immunity \cite{Regoes}. During the first
stage of HIV infection, the viruses moderately decrease the amount
of CD4$^{+}$ T cells within the host. Then, the level of CD4$^{+}$ T
cell remains almost constant for several years due to the inhibition
provided by the immune response. \cite{Komarova} modeled the
interrelationship among timing, efficiency and success of antiviral
drug therapy.  \cite{Regoes} performed investigations on a variety
of HIV models with immune impairment. The authors show that when the
impairment rate of HIV overwhelms the threshold value, immune system
of the host may collapse.  \cite{Iwami1,Iwami2} constructed
mathematical models to study the infection of HIV and carried out
analysis to obtain a `risky threshold' and  an `immunodeficiency
threshold' for the impairment rate. Their investigations implies
that when the impairment rate is greater than a threshold value, the
immune system of the host always collapses.

During the early stage of an HIV infection, latent HIV reservoirs
will be formed in the host. Latent reservoirs survive the
antiretroviral therapy (ART) and remain alive even when the level of
HIV in the blood is undetectable. Thus, the existence of such
reservoir is a barrier to the elimination of HIV. HIV infection
dynamics with latent reservoirs have been investigated in the
literature\citep{Kim1,Rong1,Rong2,Rong3}. \cite{Rong3} investigated
the HIV infection with latent reservoirs by constructing a
stochastic model. The authors showed that the latent reservoir has
relatively stable size and cells can be activated to produce
virions.  \cite{Kim1} investigated influence of ongoing viral
replication on the evolution of latent reservoirs and revealed the
influences of a variety of viral and host factors on the dynamics of
the latent reservoirs. \cite{Rong2} established a mathematical model
to investigate the activation of latently infected cells and
revealed the mechanism behind the replenishment of the latent
reservoirs.

In this article, we integrate oxidative stress into HIV infection
model to consider the interplay between viruses and corresponding
oxidative stress, and their combined effects on the host. The within
host model is given by
\begin{equation}\label{e1}
\left\{\begin{split}
& \frac{dx(t)}{dt}=s-dx(t)-(1-\epsilon)\beta x(t)y(t), \\
& \frac{dL(t)}{dt}=\alpha_{L}(1-\epsilon)\beta x(t)y(t)+(\rho-a-d_{L})L(t), \\
& \frac{dy(t)}{dt}=(1-\alpha_{L})(1-\epsilon)\beta x(t)y(t)+aL(t)-\delta y(t)-py(t)z(t), \\
& \frac{dz(t)}{dt}=\frac{cy(t)z(t)}{1+\eta y(t)}K(y)-bz(t),
 \end{split}\right.
\end{equation}
where $x$ denotes activated CD4$^{+}$ T cells, $L$ viral latent
reservoir, $y$ infected CD4$^{+}$ T cells and $z$ immune cells.
Since HIV dynamics are known to be more rapid than infected cell
dynamics, we make the quasi-steady assumption. Thus, the HIV cells
are in proportion to the infected cells. Here we use the overall
treatment effectiveness, denoted by $\epsilon,$ for
$0\leq\epsilon\leq1,$ to describe their combined treatment
effectiveness. In particular, when $\epsilon=1$, the therapy is
$100\%$ effective. On the other hand, if the treatment is
terminated,  $\epsilon=0$ \citep{Rong1,Conway}.

The relationship between the ROS and antioxidant can be evaluated
using the method proposed in \cite{Gaalen}.
\citep{Zhang1,Zhang2,Zhang3} investigated the role of ROS in HIV
infection. In this article, we are particularly interested in the
effects of the oxidative stress on the process of viral infection.
Since oxidative stress slows down the activation of the immune
system, we hereby introduce the expression
$K(y)=\frac{1+h}{\beta(y)}$ to model the influence of the oxidative
stress.  Since reactive oxygen species (ROS) are responsible for the
oxidative stress, here use $ \rho (y) = k_0 + \frac{yk_1}{y+y_0}$ to
model oxidative stress in the HIV infection model. Here, $k_0$,
$k_1$ and $y_0$ are constants. We notice that $\rho(y)$ is a
saturating, increasing function of  $y$. In order to simplify the
analysis, we linearize the expression of $\rho(y)$ to obtain
$\rho(y)= {k _0}+{\frac {{k_1}\,y}{{y_0}}} $. Letting $k_0=k$ and
${\frac {{k_1}\,y}{{y_0}}} =r$, we obtain $\rho(y)= {k }+ry $. It
thus follows that
$$K(y)=\frac{1+h}{k+ry}.$$

 ROS damage, the immune term is modeled using the expression
$\frac{cyz}{1+\eta y}K(y)-bz$. We notice that such immune term is
different from the widely used immune and immune impairment function
$\frac{cyz}{1+\eta y}-bz-myz$ \citep{De,Komarova,zwang,Pugliese} ($
m $ is the rate of immune impairment).
Here constant $h$ represents the influence of antioxidant on the
immune response, $k$ is the influence of  naturally produced
oxidant, and $r$ is the influence of oxidant produced by infected
CD4$^{+}$ T cells or HIV viral load $y$.

The rest of this article is organized as follows. In, Section 2, we
present some preliminary results on the structure of the equilibria
of the model. We then, in Section 3, perform stability analysis on
the equilibria. In Section 4, based on our stability analysis, we
present sensitive analysis and numerical simulations. Finally, in
Section 5, we conclude the paper with discussions and a summary.

\section{Preparation }

 \subsection{ Positiveness and boundedness }

In the following, we show that  system (\ref{e1}) is  well-posed.

 \begin{Theorem}\label{th21}  System (\ref{e1}) has a unique and
nonnegative solution with the initial condition
$(x(0),L(0),y(0),z(0))\in \mathbb{R}_{+}^{4},$ where
$\mathbb{R}_{+}^{4}= \{(x_{1}, x_{2}, x_{3}, x_{4})|x_{j}\geq 0, j =
1, 2, 3, 4\}.$ Furthermore, the solution is  bounded.
\end{Theorem}

\noindent{\bf Proof.} By the fundamental theory of ordinary
differential equations,   system (\ref{e1}), with nonnegative
initial conditions, has  a unique solution.
For any nonnegative initial data, let $t_{1}>0$ be the first time
that $x(t_{1})=0$.
The first equation of (\ref{e1}) implies that  $\dot{x}(t_{1})=s>0$.
That is to say,  $x(t)<0$ for $t\in(t_{1}-\varepsilon_{1}, t_{1})$,
where $\varepsilon_{1}$ is an arbitrarily small positive constant.

The above discussion leads to a contradiction. It thus follows that
$x(t)$ is always positive. Because $z=0$ is a constant solution of
the last equation of (\ref{e1}), it follows from the  fundamental
existence and uniqueness theorem that  $z>0$ for all $t>0.$

Suppose that at time $t_{2}
> 0$, $y(t_{2})z(t_{2})$  reaches $0$ for the first time. Thus, we have

(i) $L(t_{2})=0, y(t)\geq 0 $ for $t\in[0,t_{2}],$ or

(ii) $y(t_{2})=0, L(t)\geq0$ for $t\in[0, t_{2}].$

  For case(i),   because $x(t)$ is positive, it follows from the variation of constants formula
  that $L(t_{2})=L(0)+e^{-\int_{0}^{t_{2}}(a+d_{L}-\rho)d\xi}\int_{0}^{t_{2}}\alpha_{L}(1-\epsilon)\beta x(\xi)y(\xi)d\xi>0,$
 which is in contradiction with $L(t_{2})=0$.

For case (ii),  the third equation of system (\ref{e1}) implies that

$y(t_{2})=y(0)+e^{\int_{0}^{t_{2}}[(1-\alpha_{L})(1-\epsilon)\beta
x(\xi)-\delta-pz(\xi)]d\xi}\int_{0}^{t_{2}}aL(\xi)d\xi>0,$ which
contradicts $y(t_{2})=0$. Thus, $L(t)~~ \mbox{and} ~~ y(t)$ are
always positive.

Next, we expatiate upon the boundedness of the solutions to
(\ref{e1}). Let
$$ M(t)=\sigma x(t)+aL(t)+(a+d_{L}-\rho)y(t)+\frac{pk(a+d_{L}-\rho)z(t)}{c(1+h)},$$
where $\sigma=a\alpha_{L}+(1-\alpha_{L})(a+d_{L}-\rho)$. Since all
solutions of (\ref{e1}) are positive, we have
$$\begin
{array}{lll}\frac{d M }{dt}
&=&\sigma\Big[s-dx-(1-\epsilon)\beta xy\Big]+a\Big[\alpha_{L}(1-\epsilon)\beta xy+(\rho-a-d_{L})L\Big]\\
& &+(a+d_{L}-\rho)\Big[(1-\alpha_{L})(1-\epsilon)\beta xy+aL-\delta y-pyz\Big]\\
& &+\frac{pk(a+d_{L}-\rho)}{c(1+h)}\Big(\frac{cyz}{1+\eta
y}\frac{1+h}{k+ry}-bz\Big)\\
&\leq&\sigma s-\sigma dx-(a+d_{L}-\rho)\delta
y-\frac{bpk(a+d_{L}-\rho)z}{c(1+h)}\\
&<&\sigma s-\nu M.
\end {array}$$
 Here  $\nu=\min\Big\{d, \delta, b
\Big\}$. Let $\varphi$ be the solution of
$$\left\{\begin
 {array}{l l} \displaystyle \frac{d\varphi}{dt}=\sigma s-\nu\varphi, \\
 \displaystyle \varphi_{0}=\sigma x_{0}+aL_{0}+(a+d_{L}-\rho)y_{0}+\frac{pk(a+d_{L}-\rho)z_{0}}{c(1+h)},
  \end {array}
  \right. $$
  where $x_{0}, y_{0}$ and  $z_{0}$ are  initial values of
  system (\ref{e1}) and $\varphi_{0}= M_{0}>0.$
  We then evaluate $\lim_{t\rightarrow+\infty}\sup_{}\varphi(t)=\frac{\sigma s}{\nu}.
$ It follows from the comparison theorem \cite{Rubinstein} that $
M(t) <\varphi(t).$ Thus, $x(t)$, $L(t)$, $y(t)$ and $z(t)$ are
bounded. \qed

 \subsection{ Thresholds }

In the following, we consider the threshold values of the model.
Such threshold characterises the  viral dynamics of   model
(\ref{e1}).

Let
$$\begin
{array}{lll} R_0
&=&(1-\epsilon)\beta\Big[(1-\alpha_{L})+\frac{a\alpha_{L}}{a+d_{L}-\rho}\Big].\frac{s}{d}.\frac{1}{\delta}\vspace{0.2cm}\\
 &=&\frac{s\beta(1-\epsilon)\Big[a\alpha_{L}+(1-\alpha_{L})(a+d_{L}-\rho)\Big]}{d\delta(a+d_{L}-\rho)}.
\end {array}$$
Because $(1-\epsilon)\beta.\frac{s}{d}.\frac{1}{\delta}$ is the
basic reproductive number of the model without viral latent
reservoir, $R_0$ gives the basic reproductive number of model
(\ref{e1}), which describes the average number of newly infected
cells generated from an infected cell at the beginning of the
infectious process.

Let
$$\begin
{array}{lll} R_{\pm}^{*}
&=&\frac{s\beta(1-\epsilon)\Big[(1-\alpha_{L})+\frac{a\alpha_{L}}{a+d_{L}-\rho}\Big]}{\delta
\Big[d+\frac{b\beta(1-\epsilon)}{\frac{B\mp\sqrt{B^{2}-4b^{2}kr\eta}}{2k}}\Big]}\vspace{0.2cm}\\
 &=&\frac{2b\eta r s\beta(1-\epsilon)\Big[a\alpha_{L}+(a+d_{L}-\rho)(1-\alpha_{L})\Big]}{\delta(a+d_{L}-\rho)
\Big\{2b\eta
rd+\beta(1-\epsilon)\Big[B\pm\sqrt{B^{2}-4b^{2}kr\eta}\Big]\Big\}},
\end {array}$$
 where  $$B=c+ch-br-bk\eta.$$
Because $\frac{s\beta(1-\epsilon)}{\delta
[d+\frac{b\beta(1-\epsilon)}{c}]}$ is the basic immune reproductive
number of the model with the bilinear immune incidence $(cyz)$ and
without viral latent reservoir, $R_{\pm}^{*}$  represent the two
thresholds in addition to the basic reproductive ratio.

We also define the following thresholds
$$  h_{1} =\frac{br+b\eta k}{c}-\frac{2b\sqrt{\eta rk}}{c}-1$$ and
$$h^{*}=\frac{b r+b\eta k}{c}+\frac{2b\eta ry_{1}}{c}-1.$$
The post-treatment immune control threshold is then obtained as
$$
 h_{2}=\frac{br+b\eta k}{c}+\frac{2b\sqrt{\eta rk}}{c}-1,
$$ and the elite control threshold is given by
$$h^{**}=\frac{b r+b\eta k}{c}+\frac{bk\beta(1-\epsilon)}{cd(R_{0}-1)}+\frac{bd\eta r(R_{0}-1)}{c\beta(1-\epsilon)}-1.$$
Denote $R_{c}=1+\frac{\beta(1-\epsilon)\sqrt{\eta rk}}{dr\eta},$ we
have the following results.

 \begin{Lemma}\label{la1} $R_{0}>R_{c}\Leftrightarrow
h^{*}>h^{**}.$    \end{Lemma}

 \begin{Lemma}\label{la2}

 (i) If $1<R_{0}<R_{c},$ then $R_{+}^{*}<1,$
$h^{*}<h_{2}$ and $R_{-}^{*}>1\Leftrightarrow h>h^{**}.$

 (ii) If
$R_{0}>R_{c},$
 then $h^{*}>h_{2},$ $R_{-}^{*}>1\Leftrightarrow h>h_{2}$ and  $R_{+}^{*}>1\Leftrightarrow
 h_{2}<h<h^{*}$. \qed\end{Lemma}

 \subsection{ Equilibria }

 In the following, we consider the existing conditions of equilibria of system (\ref{e1}).

 System (\ref{e1}) always admits an uninfected equilibrium $E_{0}=(x_{0},0,0,
0),$ where $x_{0} =\frac{s}{d}$.

 (i) If  $R_0>1$,
system (\ref{e1}) also has an immune-free equilibrium $E_{1}=(x_{1},
L_{1}, y_{1}, 0), $ where
$$\begin
{array}{lll}
x_{1}&=&\frac{\delta(a+d_{L}-\rho)}{\beta(1-\epsilon)\Big[a\alpha_{L}+(1-\alpha_{L})(a+d_{L}-\rho)\Big]},\vspace{0.2cm}\\
L_{1}&=&\frac{\alpha_{L}\beta(1-\epsilon)x_{1}y_{1}}{a+d_{L}-\rho}, \vspace{0.2cm}\\
y_{1}&=&\frac{d(R_{0}-1)}{\beta(1-\epsilon)}.
\end {array}$$

(ii) If  $R_{-}^{*}>1$  and $0<h<h_{1}$ or $h>h_{2},$ equation
$\frac{cyz}{1+\eta y}\frac{1+h}{k+ry}-bz=0$ has two positive roots.

If  $R_{-}^{*}>1$  and  $h>h_{2},$ system (\ref{e1}) has  an immune
equilibrium $E_{-}^{*}=(x_{-}^{*}, L_{-}^{*}, y_{-}^{*},
z_{-}^{*})$. If  $R_{+}^{*}>1$ and  $h>h_{2},$ system (\ref{e1})
also has an immune equilibrium $E_{+}^{*}=(x_{+}^{*}, L_{+}^{*},
y_{+}^{*}, z_{+}^{*}).$
 Here
$$\begin
{array}{lll}
x_{\pm}^{*}&=& \frac{s}{d+\beta(1-\epsilon)y_{\pm}^{*}}, \vspace{0.2cm} \\
L_{\pm}^{*}&=&\frac{\alpha_{L}(1-\epsilon)\beta
x_{\pm}^{*}y_{\pm}^{*}}{a+d_{L}-\rho},  \vspace{0.2cm}\\
y_{\pm}^{*}&=&\frac{B\mp\sqrt{B^{2}-4b^{2}kr\eta}}{2b\eta r}, \vspace{0.2cm} \\
z_{\pm}^{*}&=&\frac{\delta(R_{\pm}^{*}-1)}{p}.\end {array}$$

From Lemmas \ref{la1} and \ref{la2}, summing up the above analysis
yields the existing results of equilibria of system (\ref{e1})

 \begin{Theorem}

(i) System (\ref{e1}) always admits an uninfected equilibrium
$E_{0}.$

(ii) If  $R_0>1$, system (\ref{e1}) also has an immune-free
equilibrium $E_{1}.$

(iii) If $1<R_{0}<R_{c}$ and  $h>h^{**},$ system (\ref{e1})  has
only one positive equilibrium $E_{+}^{*}.$

 If $R_{0}>R_{c}$ and
 $h_{2}<h<h^{*}$, system (\ref{e1})  has  two positive equilibria
$E_{-}^{*}$ and  $E_{+}^{*}$.

 When  $R_{0}>R_{c}$ and
 $h>h^{*}$, system (\ref{e1})   has only one positive equilibrium
$E_{+}^{*}$.

(iv) If $R_{0}>R_{c}$  and $h=h_{2},$ system (\ref{e1})  has only
one positive equilibrium $E_{*}.$ \qed \end{Theorem}

The existence results for positive equilibria are summarized   in
Tables 2.1 and  2.2.

\begin{table}[!h] \label{ta1}
\begin{center}
\begin{tabular}{|l|l|l|}
\hline   &  $h_{2}<h<h^{**}$  &  $h>h^{**}$

\\\hline
$E_{+}^{*}$      &\hspace{0.6cm}--- & exist  \\
$E_{-}^{*}$      &\hspace{0.6cm}--- &\hspace{0.1cm} --- \\
\hline
\end{tabular}
\end{center}
\caption{The existence of the positive equilibria when
$1<R_{0}<R_{c}.$}
\end{table}

\begin{table}[!h] \label{ta2}
\begin{center}
\begin{tabular}{|l|l|l|}
\hline   &  $h_{2}<h<h^{*}$  &  $h>h^{*}$

\\\hline
$E_{+}^{*}$      &\hspace{0.5cm}exist & exist  \\
$E_{-}^{*}$      &\hspace{0.5cm}exist &\hspace{0.1cm} --- \\
\hline
\end{tabular}
\end{center}
\caption{The existence of the positive equilibria when
$R_{0}>R_{c}.$}
\end{table}

\section{Stability analysis}

In this section, we  consider the stabilities of equilibria for
system (\ref{e1}).

Let $\tilde{E}$ be any arbitrary equilibrium of system (\ref{e1}).
Denote
$$\mathscr{J}=\left[
\begin{array}{cccc}
  -d-\beta(1-\epsilon)\tilde{y} & 0 & -\beta(1-\epsilon)\tilde{x} & 0\\
  \alpha_{L}\beta(1-\epsilon)\tilde{y} & \rho-a-d_{L} & \alpha_{L}(1-\epsilon)\beta \tilde{x}  & 0 \\
  (1-\alpha_{L})\beta(1-\epsilon)\tilde{y} & a & (1-\alpha_{L})\beta(1-\epsilon)\tilde{x}-\delta-p\tilde{z} & -p\tilde{y}\\
  0 & 0 &
  \frac{c(1+h)\tilde{z}(k-\eta r\tilde{y}^{2})}{(1+\eta\tilde{y})^{2}(k +r\tilde{y})^{2}}  & \frac{c(1+h)\tilde{y}}{(1+\eta\tilde{y})(k+r\tilde{y})}-b \\
\end{array}
\right].$$ The characteristic equation of the linearized system of
(\ref{e1}) at $\tilde{E}$ is then obtained as
$$  \left|\lambda I-\mathscr{J} \right|=0.
 \eqno(3.1)$$

\subsection{Stability analysis of Equilibrium $E_{0}$}

 \begin{Theorem}\label{th31}£¬  If $R_0<1$, then the uninfected
equilibrium $E_{0}$ of system (\ref{e1}) is locally  asymptotically
stable. If $R_0>1$,  $E_{0}$ is unstable.  \end{Theorem}

\noindent{\bf Proof.}  The characteristic equation (3.1) with
respect to  equilibrium $E_{0}(x_{0}, 0, 0, 0)$  is $$  \left|
\begin{array}{cccc}
  -d-\lambda & 0 & -\beta(1-\epsilon)x_{0} & 0\\
  0 & \rho-a-d_{L}-\lambda & \alpha_{L}(1-\epsilon)\beta x_{0}  & 0 \\
  0 & a & (1-\alpha_{L})(1-\epsilon)\beta x_{0}-\delta-\lambda  & 0\\
  0 & 0 &
  0 & -b-\lambda \\
\end{array}
\right|=0. \eqno(3.2)$$
 It is clear that   equation (3.2) has two
negative roots $ -d$ and $-b.$    The other two eigenvalues are
solutions of

$$
 \lambda^{2}+a_{1}\lambda+a_{2}=0,
 \eqno(3.3)$$
where
\begin{equation}\nonumber
\begin{split}
a_{1}
&=a+d_{L}-\rho+\delta[1-\frac{(1-\alpha_{L})(1-\epsilon)\beta x_{0}}{\delta}],  \vspace{0.2cm}\\
a_{2}&=(a+d_{L}-\rho)-a\beta(1-\epsilon)[\delta-(1-\alpha_{L})(1-\epsilon)\beta x_{0}]-\frac{as\beta\alpha_{L}(1-\epsilon)}{d} \vspace{0.2cm}\\
&=\delta(a+d_{L}-\rho)(1-R_{0}).
\end{split}
\end{equation}

It is easy to see that $a_{1}>0$ and $a_{2}>0$ for $ R_{0}<1$.  When
$ R_{0}<1$,
 equation (3.3) has  two
negative roots indicating that  $E_{0}$ is locally stable.  On the
other hand, when $R_{0}>1$,  then $a_{2}<0$, and $E_{0}$ is a saddle
with dim $W^{s}(E_{0})=2$ and dim$W^{u}(E_{0})=1$, and hence
unstable.  This completes the proof of Theorem \ref{th31}.
   \qed

 \begin{Theorem} If $R_0<1$, then the uninfected
equilibrium $E_{0}$ of system (\ref{e1}) is global  asymptotically
stable.  \end{Theorem}

\noindent{\bf Proof.} Define a function
$$V=\frac{1}{2}(x-x_{0})^{2}+AL+By+\frac{ pB}{c(1+h)}z,$$
where $A$ and $B$ are  positive coefficients to be undetermined. It
is easy to see that $V$ is a positive Lyapunov function. Evaluating
the time derivative of $V$ along the solution of system (\ref{e1})
yields
\begin{eqnarray*}
\dot{V}|_{(\ref{e1})}&=&(x-x_{0})\Big[s-dx-(1-\epsilon)\beta xy\Big]+A\Big[\alpha_{L}(1-\epsilon)\beta xy-(a+d_{L}-\rho)L\Big] \vspace{0.3cm}\\
& &+B\Big[(1-\alpha_{L})(1-\epsilon)\beta xy+aL-\delta
y-pyz\Big]+\frac{ pB}{c(1+h)}\Bigg(\frac{cyz}{1+\eta
y}\frac{1+h}{k+ry}-bz\Bigg)  \vspace{0.3cm}\\
&=&(x-x_{0})\Big[dx_{0}-dx-(1-\epsilon)\beta xy+(1-\epsilon)\beta x_{0}y-(1-\epsilon)\beta x_{0}y\Big]  \vspace{0.3cm}\\
& & +A\alpha_{L}(1-\epsilon)\beta xy-A(a+d_{L}-\rho)L+B(1-\alpha_{L})(1-\epsilon)\beta xy  \vspace{0.3cm}\\
& &+BaL-B\delta y-Bpyz+\frac{ pB}{c(1+h)}\frac{cyz}{1+\eta
y}\frac{1+h}{k+ry}-\frac{ pB}{c(1+h)}bz \vspace{0.3cm}\\
&\leq&-\Big(d+(1-\epsilon)\beta
y\Big)(x-x_{0})^2-\Big[x_{0}-A\alpha_{L}-B(1-\alpha_{L})\Big](1-\epsilon)\beta
xy  \vspace{0.3cm}\\
& &-\Big[B\delta-(1-\epsilon)\beta x_{0}^{2}\Big]y
-\Big[A(a+d_{L}-\rho)-Ba\Big]L-\frac{ pB}{c(1+h)}bz.
\end{eqnarray*}
 Choosing
\begin{eqnarray*}
A&=&\frac{x_{0}}{(1-\alpha_{L})[\frac{a+d_{L}-\rho}{a}+\frac{\alpha_{L}}{1-\alpha_{L}}]},\vspace{0.3cm}\\
B&=&\frac{A(a+d_{L}-\rho)}{a},
\end{eqnarray*}
we get  \begin{eqnarray*}
x_{0}-A\alpha_{L}-B(1-\alpha_{L})&\geq&0,\\
B\delta-(1-\epsilon)\beta x_{0}^{2}&\geq&0,\\
A(a+d_{L}-\rho)-Ba&\geq&0.
\end{eqnarray*}
Thus,  if $R_0\leq1,$ we have $\dot{V}|_{(\ref{e1})}\leq0.$ Since
$x, L, y, z$ are positive, we get $\dot{V}=0$ if and only if $(x, L,
y, z)=(x_{0}, 0, 0)$. It thus follows from the  classical
Krasovskii-LaSalle principle \citep{Krasovskii,LaSalle} that $E_{0}$
is globally asymptotically stable.           \qed

The global asymptotic stability of the uninfected equilibrium
$E_{0}$ of system (\ref{e1}) biologically implies that the virus
will die out in the host. Generally, with treatment  strong enough,
we have $R_0<1$ which guarantees the elimination of the virus.

\subsection{Stability analysis of Equilibrium $E_{1}$}

 \begin{Theorem}\label{th33} Assume $R_0>1$,  if $h<h_{1}$,
$h_{1}<h<h_{2}$ or $h_{2}<h<h^{**},$ then the immune free
equilibrium $E_{1}$ of system (\ref{e1}) is locally asymptotically
stable. If $h>h^{**} ,$ $E_{1}$ is unstable.
\end{Theorem}

\noindent{\bf Proof.}  The characteristic equation of the linearized
system of (\ref{e1}) at $E_{1}$ is given by
$$(\lambda^{3}+b_{1}\lambda^{2}+b_{2}\lambda+b_{3})\Bigg[\frac{c(1+h)y_{1}}{(1+\eta
y_{1})(k+ry_{1})}-b\Bigg]=0,$$ where $$\begin {array}{lll}
b_{1} &=&d+(1-\epsilon)\beta y_{1}+\underbrace{a+d_{L}-\rho}_{\textcircled{\small{1}}}+\underbrace{\frac{a\alpha_{L}(1-\epsilon)\beta x_{1}}{a+d_{L}-\rho}}_{\textcircled{\small{2}}}\\
b_{2}&=&d(a+d_{L}-\rho+\frac{aL_{1}}{y_{1}}) +(1-\epsilon)\beta
aL_{1}+\underbrace{(1-\epsilon)\beta y_{1}(a+d_{L}-\rho)}_{\textcircled{\small{3}}}\\
& & +\underbrace{(1-\epsilon)\beta x_{1}(1-\alpha_{L})(1-\epsilon) \beta y_{1}}_{\textcircled{\small{4}}}\\
b_{3}&=&a\alpha_{L}(1-\epsilon)\beta x_{1}(1-\epsilon)\beta
y_{1}+(a+d_{L}-\rho)(1-\epsilon)\beta
x_{1}(1-a_{L})(1-\epsilon)\beta y_{1}.
\end {array}$$
It is easy to see that
$$\textcircled{\small{1}}\times\textcircled{\small{4}}+\textcircled{\small{2}}\times\textcircled{\small{3}}-b_{3}=0.$$
Thus, $b_{1}b_{2}-b_{3}>0$ holds true. Now, we discuss the sign of
the eigenvalue
$$\begin {array}{lll}
\displaystyle \lambda_{4} &=&\displaystyle
\frac{c(1+h)y_{1}}{(1+\eta
y_{1})(k+ry_{1})}-b \vspace{0.2cm}\\
&=&\displaystyle \frac{-br\eta
y_{1}^{2}+(c+ch-br-bk\eta)y_{1}-bk}{(1+\eta y_{1})(k+ry_{1})},
\vspace{0.2cm}
\end {array}$$
which is  determined by
$$\Delta=(c+ch-br-bk\eta)^{2}-4b^{2}kr\eta.$$

(i) If $\Delta=0,$ then $h=h_{1}$ or $h=h_{2},$ which is a critical
situation.

(ii) If $\Delta<0,$ then $h_{1}<h<h_{2},$ we have $\lambda_{4}<0.$

%%%
(iii) If $\Delta>0,$ we have $h<h_{1}$ or $h>h_{2}.$ To get
$\lambda_{4}<0,$ we need to ensure that $h<\frac{br+b\eta k}{c}-1$,
$R_{0}<1+R_{1}$ or $R_{0}>1+R_{2},$ from which we can obtain that
$h<h^{**}.$ %%
Here $R_{1,2}=\frac{\beta(1-\epsilon)\Big[B\mp\sqrt{B^2-4b^{2}\eta
rk}\Big]}{2bdr\eta}.$ Notice $h_{2}<h^{**}.$ It thus follows that if
$h<\frac{br+b\eta k}{c}-1$ or $h_{2}<h<h^{**},$ then the eigenvalue
$\lambda_{4}<0.$ If $h>h^{**},$ we have $\lambda_{4}>0.$

In summary, if $h<h_{2}$ or $h_{2}<h<h^{**},$ then $\lambda_{4}<0.$
From the  Routh-Hurwitz criterion \cite{Routh,Hurwitz}, with the
assumption
  $R_{0}>1$, if $h<h_{2}$ or $h_{2}<h<h^{**},$ the equilibrium
$E_{1}$ of system (\ref{e1}) is locally asymptotically stable. On
the other hand, when $h>h^{**},$ $E_{1}$ is unstable. \qed

  \begin{Remark}  (i) $h_{1}, h_{2} ~\mbox{and}~  h^{**}$ are
 critical values.

 (ii) If $R_0>1$ and
$h>h^{**}$, then the equilibrium $E_{1}$ of system (\ref{e1}) is
unstable. \end{Remark}

Here, the elite control threshold $h^{**}$ determines whether a
system is under  elite control \cite{Conway}. Biologically, if the
proliferation rate of CTLs is  greater than the critical value
$h^{**}$, the virus may remain at high levels
 with no control.

\subsection{Stability analysis of positive equilibria}

We denote by  $E^{*}=(x^{*}, L^{*}, y^{*}, z^{*})$  an arbitrary
positive equilibrium of system (\ref{e1}).

 \begin{Theorem}\label{th34}

(i)  Assume  $~( \mathbf{A})~ A_{3}(A_{1}A_{2}-A_{3})-
A_{1}^{2}A_{4}>0.$  If

\hspace{0.4cm} ($\mathbf{A.1}$) \hspace{0.2cm}    $1<R_{0}<R_{c}$
and $h>h^{**}$, or

\hspace{0.4cm} ($\mathbf{A.2}$) \hspace{0.2cm}    $R_{0}>R_{c}$ and $h>h_{2}$, \\
system (\ref{e1}) has  an immune equilibrium $E_{+}^{*},$ which is a
stable node.

(ii) If  $R_{0}>R_{c}$ and
 $h_{2}<h<h^{*}$, system (\ref{e1})
also has an immune equilibrium $E_{-}^{*},$ which is an unstable
saddle point.
 \end{Theorem}

\noindent{\bf Proof.} The characteristic equation of the linearized
system of (\ref{e1}) at the arbitrary positive equilibrium $E^{*}$
is obtained as
$$\lambda^{4}+A_{1}\lambda^{3}+A_{2}\lambda^{2}+A_{3}\lambda+A_{4}=0,$$
where $$\begin {array}{lll}
A_{1} &=&a+d_{L}-\rho+d+\beta (1-\epsilon)y^{*}+\frac{aL^{*}}{y^{*}}, \vspace{0.3cm}\\
A_{2}&=&(a+d_{L}-\rho)\Big[d+\beta (1-\epsilon)y^{*}\Big]+\frac{aL^{*}}{y^{*}}\Big[d+\beta (1-\epsilon)y^{*}\Big]  \vspace{0.3cm}\\
& &+py^{*}z^{*}\frac{c(1+h)z^{*}(k-\eta ry^{*2})}{(1+\eta
y^{*})^{2}(k
 +ry^{*})^{2}}  +(1-\alpha_{L})(1-\epsilon)\beta x^{*}(1-\epsilon)\beta y^{*},  \vspace{0.3cm}\\
A_{3}&=& \frac{aL^{*}}{y^{*}}(a+d_{L}-\rho)(1-\epsilon)\beta
y^{*}+py^{*}z^{*}\frac{c(1+h)z^{*}(k-\eta ry^{*2})}{(1+\eta
y^{*})^{2}(k
 +ry^{*})^{2}}  \Big[a+d_{L}-\rho+d+\beta (1-\epsilon)y^{*}\Big] \vspace{0.3cm} \\
   & &+(1-\alpha_{L})(1-\epsilon)\beta x^{*}(1-\epsilon)\beta y^{*}(a+d_{L}-\rho),  \vspace{0.3cm}\\
A_{4}&=& py^{*}z^{*}\frac{c(1+h)z^{*}(k-\eta ry^{*2})}{(1+\eta
y^{*})^{2}(k
 +ry^{*})^{2}} (a+d_{L}-\rho)\Big[d+\beta
(1-\epsilon)y^{*}\Big].
\end {array}$$
Then we have
$$\begin {array}{lll}
A_{1}A_{2}-A_{3}&=&\frac{aL^{*}}{y^{*}}d(a+d_{L}-\rho)+(\frac{aL^{*}}{y^{*}})^{2}\Big[d+\beta (1-\epsilon)y^{*}\Big]  \vspace{0.3cm}\\
& &+\frac{aL^{*}}{y^{*}}py^{*}z^{*}\frac{c(1+h)z^{*}(k-\eta
ry^{*2})}{(1+\eta y^{*})^{2}(k
 +ry^{*})^{2}}  +\frac{aL^{*}}{y^{*}}(1-\alpha_{L})(1-\epsilon)\beta x^{*}(1-\epsilon)\beta y^{*}  \vspace{0.3cm}\\
& &+\Big[a+d_{L}-\rho+d+\beta
(1-\epsilon)y^{*}\Big](a+d_{L}-\rho)\Big[d+\beta
(1-\epsilon)y^{*}\Big]  \vspace{0.3cm}\\
& &+\frac{aL^{*}}{y^{*}}\Big[d+\beta
(1-\epsilon)y^{*}\Big]\Big[a+d_{L}-\rho+d+\beta (1-\epsilon)y^{*}\Big]  \vspace{0.3cm}\\
& &+(1-\alpha_{L})(1-\epsilon)\beta x^{*}(1-\epsilon)\beta
y^{*}\Big[a+d_{L}-\rho+d+\beta (1-\epsilon)y^{*}\Big].
\end {array}$$

(i) For equilibrium $E_{+}^{*},$ we have $$ k-\eta
ry_{-}^{*2}<0\Leftrightarrow h>h_{2}. $$

If $h>h_{2},$  then $A_{4}<0.$ Clearly, $A_{i}>0, i=1, 2, 3$ and
$A_{1}A_{2}-A_{3}>0.$  If $A_{3}(A_{1}A_{2}-A_{3})-
A_{1}^{2}A_{4}>0,$  from the Routh-Hurwitz criterion
\cite{Routh,Hurwitz}, we know that the positive equilibrium
$E_{+}^{*}$ is a stable node.

(ii) For equilibrium $E_{-}^{*},$ we have $$ k-\eta
ry_{+}^{*2}>0\Leftrightarrow
B-2b\sqrt{rk\eta}<\sqrt{B^{2}-4b^{2}rk\eta}. \eqno(3.4)$$

For any positive $h$,  (3.4) holds true. Thus, equilibrium
$E_{-}^{*}$ is unstable. \qed

 By Theorems \ref{th33} and \ref{th34}, we have the following
 result.

 \begin{Theorem} If $R_{0}>R_{c}$ and
 $h=h_{2}$, the immune equilibrium  $E_{-}^{*}$ and  $E_{+}^{*}$ coincide with each other and a saddle-node bifurcation occurs
  when $h$ passes through $h_{2}$. \qed \end{Theorem}

The stabilities of the equilibria and the behaviors of system (1.1)
are summarized  in Tables 3 and 4.
\begin{table*}[ht]
\caption{The stabilities of the equilibria and the behaviors of
system (1.1). Here, $h^{**}$ is a critical value and we assume
$A_{3}(A_{1}A_{2}-A_{3})- A_{1}^{2}A_{4}>0.$}
\begin{center}
\begin{tabular}{|l|lllll|l|}
\hline   & $E_{0}$  &$E_{1}$
 & $E_{+}^{*}$ &  $E_{-}^{*}$  &System (1.1)
\\\hline
$R_{0}<1$                        &GAS &--- & ---  & ---& Tends to $E_{0}$ \\
\hline
 $1<R_{0}<R_{c},$  $0<h<h^{**}$     &US & LAS & ---  & ---  & Tends to $E_{1}$  \\
 $1<R_{0}<R_{c},$ $h^{**}<h$     & US &  US  & LAS   &  --- & Tends to $E_{+}^{*}$ \\
\hline
\end{tabular}
\end{center}
\end{table*}
\begin{table*}[ht]
\caption{The stabilities of the equilibria and the behaviors of
system (1.1). Here, $h_{2}, h^{*} ~\mbox{and}~ h^{**}$ are critical
values, $h_{2}$ is a saddle-node bifurcation point and we assume
$A_{3}(A_{1}A_{2}-A_{3})- A_{1}^{2}A_{4}>0.$}
\begin{center}
\begin{tabular}{|l|lllll|l|}
\hline   & $E_{0}$  &$E_{1}$
 & $E_{+}^{*}$ &  $E_{-}^{*}$  &System (1.1)
\\\hline
$R_{0}<1$                        &GAS &--- & ---  & ---& Tends to $E_{0}$ \\
\hline
$R_{0}>1$, $0<h<h_{2},$         & US & LAS & --- &---& Tends to $E_{1}$\\
\hline
 $R_{0}>R_{c},$  $h_{2}<h<h^{**}$     &US & LAS & LAS  & US  & Bistable  \\
 $R_{0}>R_{c},$ $h^{**}<h<h^{*}$     & US &  US  & LAS   &  US & Tends to $E_{+}^{*}$ \\
$R_{0}>R_{c},$ $h>h^{*}$         & US & US  & LAS    &  --- &  Tends to $E_{+}^{*}$  \\
\hline
\end{tabular}
\end{center}
\end{table*}

\section{Sensitive analysis  and numerical simulations}

\subsection{Sensitive analysis}

Sensitive analysis has been widely performed to investigate the
basic reproductive number $R_{0}$ in epidemic models \cite{Xiao}. In
the following, we carry out sensitive analysis with the aim of
revealing the relationship between the basic infection reproductive
number $R_{0}$ and the basic immune reproductive number $R_{-}^{*}$,
and system parameters in our model. Here, we use latin hypercube
sampling (LHS) and partial rank correlation coefficients (PRCCs)
\citep{Blower,Marino} to test the dependence of the basic infection
reproduction number $R_{0}$ and the basic immune reproduction number
$R_{-}^{*}$. As a statistical sampling method, LHS provides an
efficient analysis of parameter variations across simultaneous
uncertainty ranges in each parameter \cite{Blower}. PRCC, on the
other hand, shows the level of significance for each parameter. The
PRCC is obtained using the rank transformed LHS matrix and output
matrix \cite{Marino}. We performed 4000 simulations per run and used
a uniform distribution function to test for the significance of
PRCCs for all parameters with wide ranges.

PRCC results Figs. 1 and 2 illustrate the dependence of $R_{0}$ and
$R_{-}^{*}$ on each system parameter respectively,  and estimate the
normal distributions for
 $R_{0}$ and $R_{-}^{*}$. When $| \mbox{PRCC} |>0.4$, there is significant correlation between input parameters and output variables. For $| \mbox{PRCC} | \in ( 0.2, 0.4]$, the correlations are moderate. When $| \mbox{PRCC} | \in [ 0, 0.2]$, we have weak correlations. We notice that the proliferation rate of CD4$^{+}$ T cells $s$, the decay rate of
CD4$^{+}$ T cells  $d$, the infection rate of CD4$^{+}$ T cells
$\beta$, the drug efficacy $\epsilon$ and the latently infected cell
death rate  $d_{L}$ have significant influences on the infection
reproduction number $R_{0} $ and  the immune reproduction number
$R_{-}^{*}$.

\subsection{Numerical simulations}

In the following,  we perform some numerical simulations to verify
our analysis results. The default parameter values are listed in
Table 5.5.

Using these default parameters,  we obtain  the values of thresholds
$R_{0}\approx3.0030$, $R_{c}\approx1.4243,$  $h_{2}\approx 0.7325$,
$h^{*}\approx 1.4353$ and $h^{**}\approx 0.9174$. The bistable
interval is $(0.7325, 0.9174).$   Fig.3 indicates that  there is no
positive equilibrium for $h<0.7325$,  and a saddle-node bifurcation
appears when $h$ passes through $0.7325$.

We are also interested in the influences of system parameters on the
virus rebound threshold $h_{2}$ and the elite control threshold $
h^{**}. $ From PRCCs. Fig.5, we can see that the decay rate of CTLs
$b$, the effector cell production Hill function scaling $\eta$, the
natural oxidant content $k$ significantly positively correlated to
the virus rebound threshold $h_{2}$. The proliferation rate of CTLs
$c$ significantly negatively correlated  to the virus rebound
threshold $h_{2}$.

 Fig.6  indicates that the activation rate of
viral latent reservoir $a$ is significantly positively correlated to
the elite control threshold $h^{**}.$  The proliferation rate
of latently infected cells $\rho$ is significantly negatively
correlated to
 the elite control threshold
$h^{**}.$

Biologically, the increased decay rate of CTLs, the effector cell
production Hill function scaling and the natural oxidant content
make it difficult to treat the disease. While the increased
proliferation rate of CTLs are beneficial to the disease treatment.

\section{ Discussion}

 The bistability phenomenon can also appear
in other HIV infection model with oxidative stress.  For example, we
investigate the HIV infection model (\ref{e51}) with logistic
proliferation rate of latently infected cells, which can reveal the
effects of proliferation rate of latently infected cells on HIV
infection model.  Instead of using similar method as analyzing
system (\ref{e1}), we carry out simulations to show the existence of
bistability.  Fig. 7 shows that system (\ref{e51}) has bistable
behaviors for different initial values when $L_{max}=50$ ( the
values of other parameters are listed in Table 2).

\begin{equation}\label{e51}
\left\{\begin{split}
&\frac{dx(t)}{dt}=s-dx(t)-(1-\epsilon)\beta x(t)y(t), \\
&\frac{dL(t)}{dt}=\alpha_{L}(1-\epsilon)\beta x(t)y(t)-(a+d_{L})L(t)+\rho L(t)(1-\frac{L(t)}{L_{max}}), \\
&\frac{dy(t)}{dt}=(1-\alpha_{L})(1-\epsilon)\beta x(t)y(t)+aL(t)-\delta y(t)-py(t)z(t), \\
&\frac{dz(t)}{dt}=\frac{cy(t)z(t)}{1+\eta
y(t)}\frac{1+h}{k+ry(t)}-bz(t).
 \end{split}\right.
\end{equation}

In fact, the function $\frac{cy}{1+\eta y}\frac{1+h}{k+ry}$ is a
Monod-Haldane function  \cite{Andrews} about $y$. We show the
predator-prey system with Monod-Haldane function or simplified
Monod-Haldane function also has bistability appear \cite{Wang-Li1}.
In viral infection systems, the models with nonmonotonic immune
responses has  bistability appear. However,  the model with
monotonic immune responses has no bistability appear
\cite{Wang-Li2}. The bistability phenomenon also be discovered in a
NK-tumor-immune system \cite{Wang-Li3}.

In this paper, we design a simplified within host model to
investigate the post-treatment immune control and elite control of a
disease.  We obtain the model's post-treatment immune control
threshold and the elite control threshold, and show that the model
displays rich dynamical behaviors. By performing sensitive analysis
and numerical simulations, we find that decreasing the immune
impairment rate is beneficial for the host to obtain post-treatment
immune control and the elite control. A therapeutic strategy that
decreases the immune impairment rate of virus, decay rate of CTLs
and effector cell production Hill function scaling is helpful for
the host to obtain elite control efficiently. The results have
potential applications in designing optimal treatment plan for
corresponding diseases.

%\begin{acknowledgements}
%If you'd like to thank anyone, place your comments here
%and remove the percent signs.
%\end{acknowledgements}

% BibTeX users please use one of
%\bibliographystyle{spbasic}      % basic style, author-year citations
%\bibliographystyle{spmpsci}      % mathematics and physical sciences
%\bibliographystyle{spphys}       % APS-like style for physics
%\bibliography{}   % name your BibTeX data base

% Non-BibTeX users please use

\begin{table}[!h]

\begin{center}
\caption{Parameters for model (1.1).}
\begin{tabular}{lllllll}
\hline Symbol Description&Value&Reference
\\ \hline
$s$  ~~~Proliferation rate of CD4$^{+}$ T cells       & 10  cells /$\mu$ L/ day &  \cite{Bonhoeffer}  \\
$d$ ~~~Decay rate of CD4$^{+}$ T cells             &0.01  day$^{-1}$&    \cite{Bonhoeffer} \\
$\beta$ ~~~Infection rate of CD4$^{+}$ T cells  & 0.015  $\mu$ L / day &  -- \\
$\epsilon$  ~~~Drug efficacy                                  & 0.8  &  --\\
$\alpha_{L}$ Fraction of newly infected cells that become latently infected    & 0.001 & --  \\
$\rho$  ~~~Proliferation rate of latently infected cells             & 0.0045 day$^{-1}$ &  \cite{Conway}   \\
$a$    ~~~Activation rate                                & 0.001 day$^{-1}$&    \cite{Conway}   \\
$d_{L}$ ~~Latently infected cell death rate                  & 0.004  day$^{-1}$&  \citep{Conway}   \\
$\delta$  ~~~Infected cell death rate                         & 1 day$^{-1}$ &  \citep{Conway}   \\
$p$ ~~~Killing rate of infected CD4$^{+}$ T cells             &0.1 day$^{-1}$&  --  \\
$c$ ~~~Proliferation rate of CTLs                             & 0.1 day$^{-1}$ &  --  \\
$\eta$ ~~Effector cell production Hill function scaling     &1  cells/$\mu$ L&  --   \\
$b$ ~~~Decay rate of CTLs                       &0.1 day$^{-1}$ &  --   \\
$h$ ~~~Antioxidant paremater             &0.8 day$^{-1}$ &  --    \\
$k$ ~~~The natural oxidant content            &1  cells /$\mu$ L &  --    \\
$r$ ~~~Oxidant content produced by  HIV viral load            &0.1
cells/$\mu$
$L$ &    --    \\
 \hline
\end{tabular}
\end{center}
\end{table}

\begin{figure}[!h]
\begin{center}
{\rotatebox{0}{\includegraphics[width=0.86 \textwidth,
height=56mm]{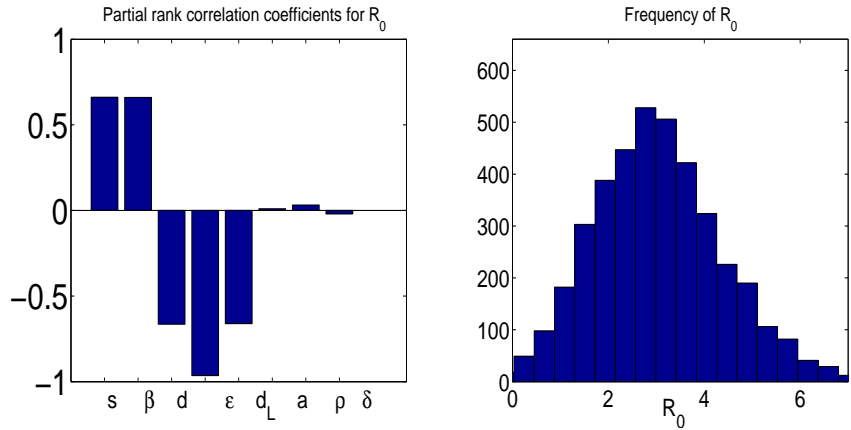}}}
 \caption{
\footnotesize  Partial rank correlation coefficients illustrating
the dependence of $R_0$  for the model (1.1) on each parameter and
the frequency distribution of $R_0.$ }\label{F51}
\end{center}
 \end{figure}

 \begin{figure}[!h]
\begin{center}
{\rotatebox{0}{\includegraphics[width=0.86 \textwidth,
height=56mm]{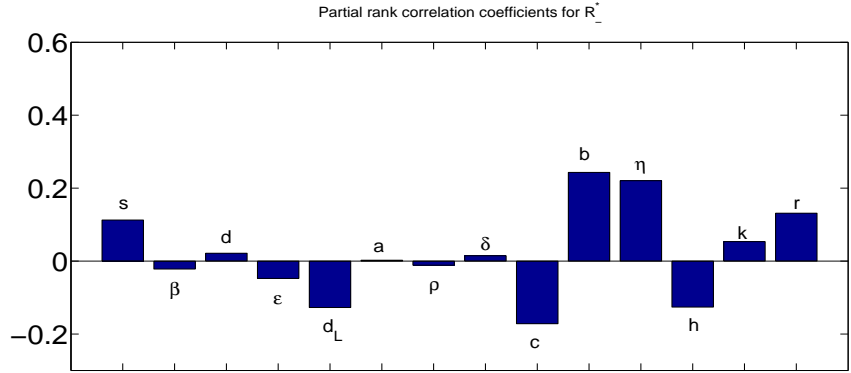}}}
 \caption{
\footnotesize Partial rank correlation coefficients illustrating the
dependence of $R_-^{*}$ for the model (1.1) on each parameter.
}\label{F51}
\end{center}
 \end{figure}

\begin{figure}[!h]
\begin{center}
{\rotatebox{0}{\includegraphics[width=0.96 \textwidth,
height=76mm]{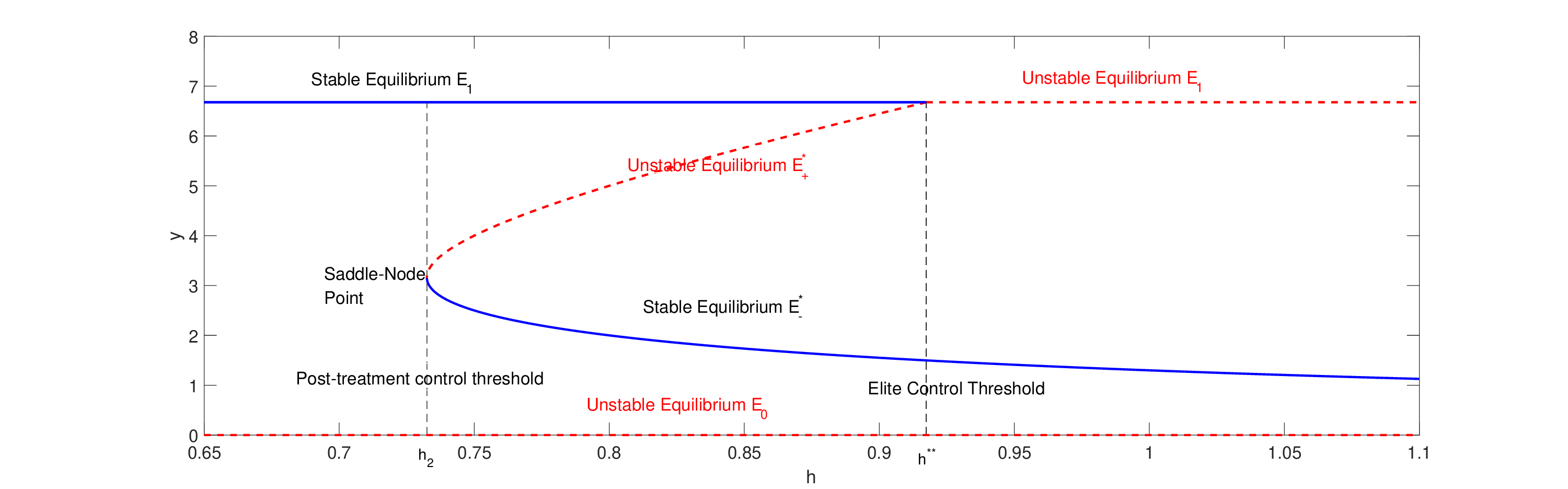}}}
 \caption{
\footnotesize Bistability and saddle-node bifurcation diagram of
system (\ref{e1}). The bistable interval is $(0.7325, 0.9174).$ When
$h<0.7325$, the model has a high viral load steady state, which
corresponds to viral rebound. When
 $h>0.9174$,  the model shows a low viral load steady state, which
means that patients are under elite control. When $h$ is between the
two values, the model shows bistability depending on the initial
conditions (population size of infected cells or immune cells at the
time of treatment cessation). The parameter values are shown in
Table 5. }\label{F51}
\end{center}
 \end{figure}

\begin{figure}[!h]
\begin{center}
{\rotatebox{0}{\includegraphics[width=0.96 \textwidth,
height=96mm]{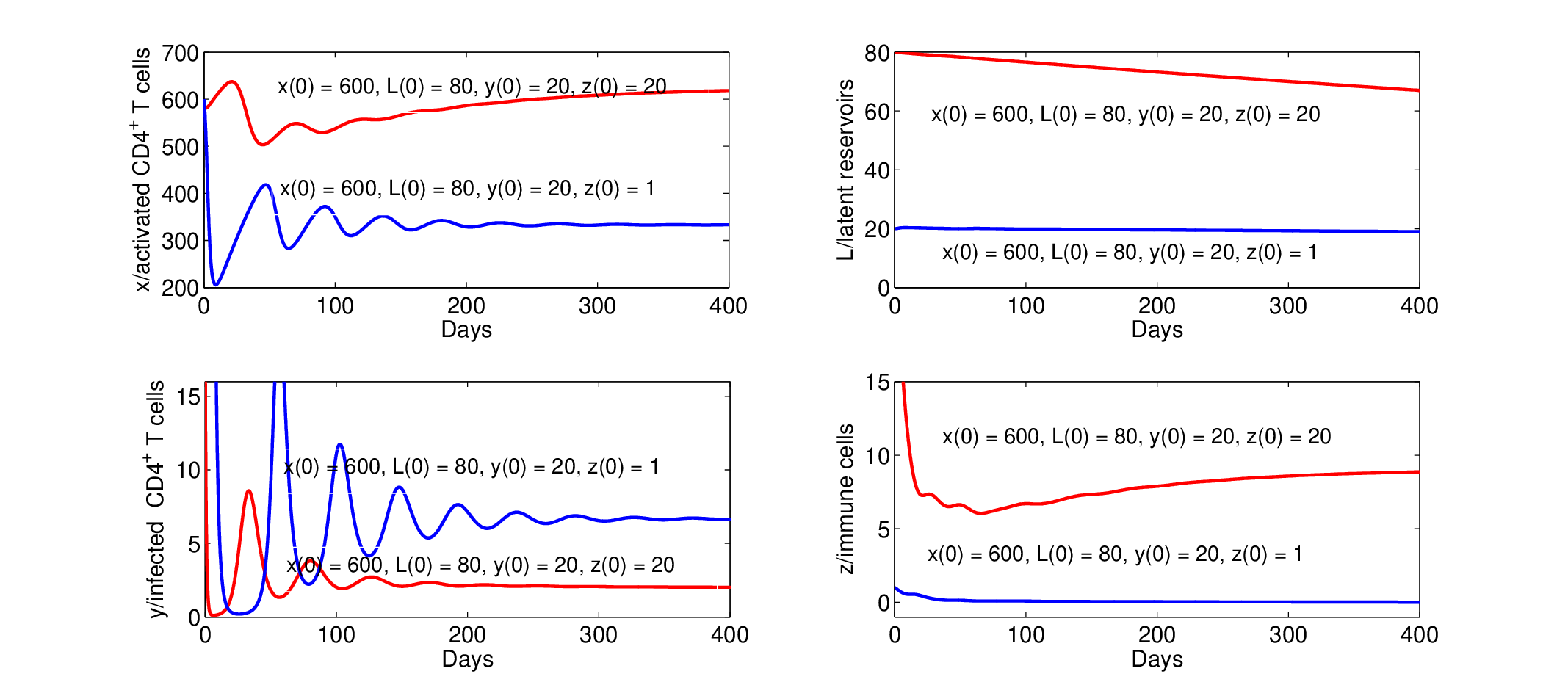}}}
 \caption{
\footnotesize   The bistability of system (\ref{e1}). The initial
values are $x(0)=600$, $L(0)=80$, $y(0)=20$, $z(0)=1$ (blue) and
$x(0)=600$, $L(0)=80$, $y(0)=20$, $z(0)=20$ (red). The parameter
values are listed in Table 5. }\label{F51}
\end{center}
 \end{figure}

  \begin{figure}[!h]
\begin{center}
{\rotatebox{0}{\includegraphics[width=0.76 \textwidth,
height=76mm]{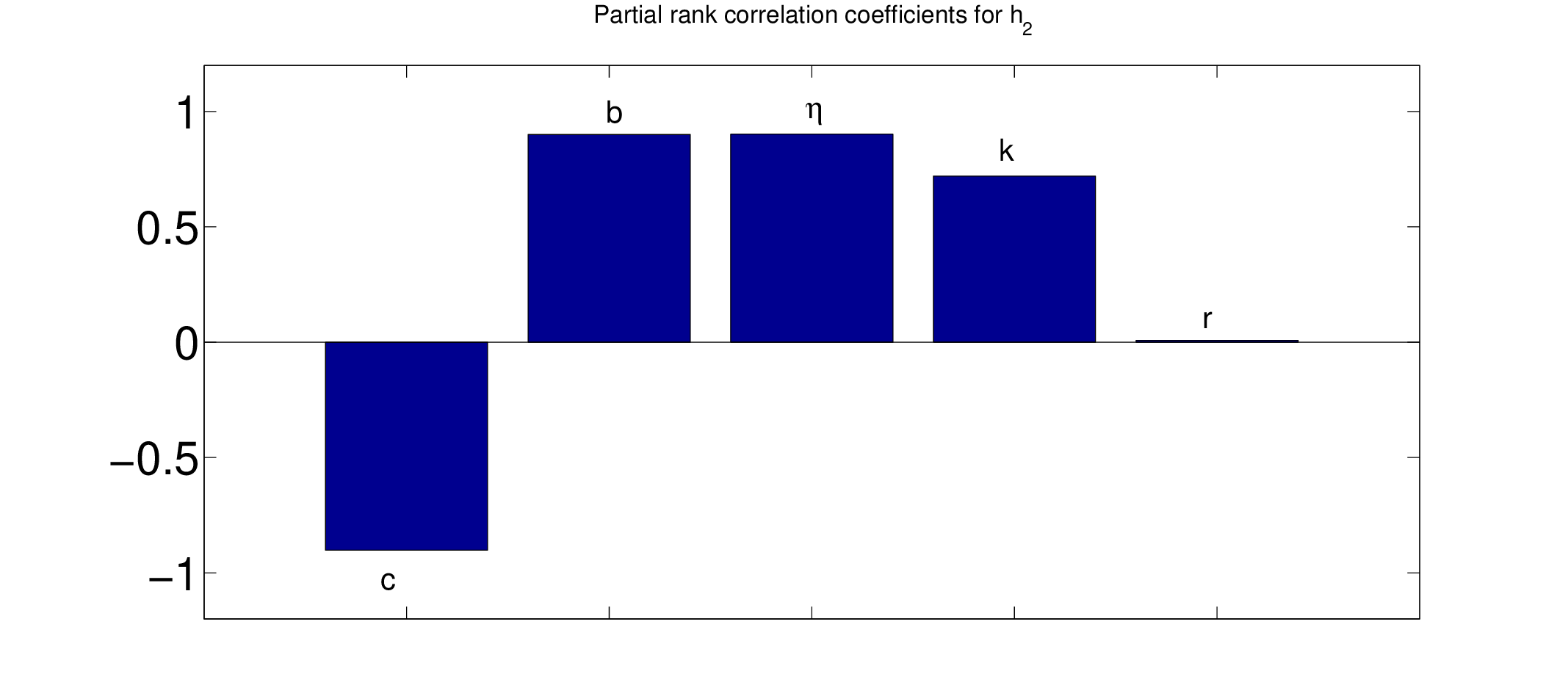}}}
 \caption{
\footnotesize  Partial rank correlation coefficients illustrating
the dependence of $h_{2}$ on each parameter. }\label{F51}
\end{center}
 \end{figure}

  \begin{figure}[!h]
\begin{center}
{\rotatebox{0}{\includegraphics[width=0.76 \textwidth,
height=76mm]{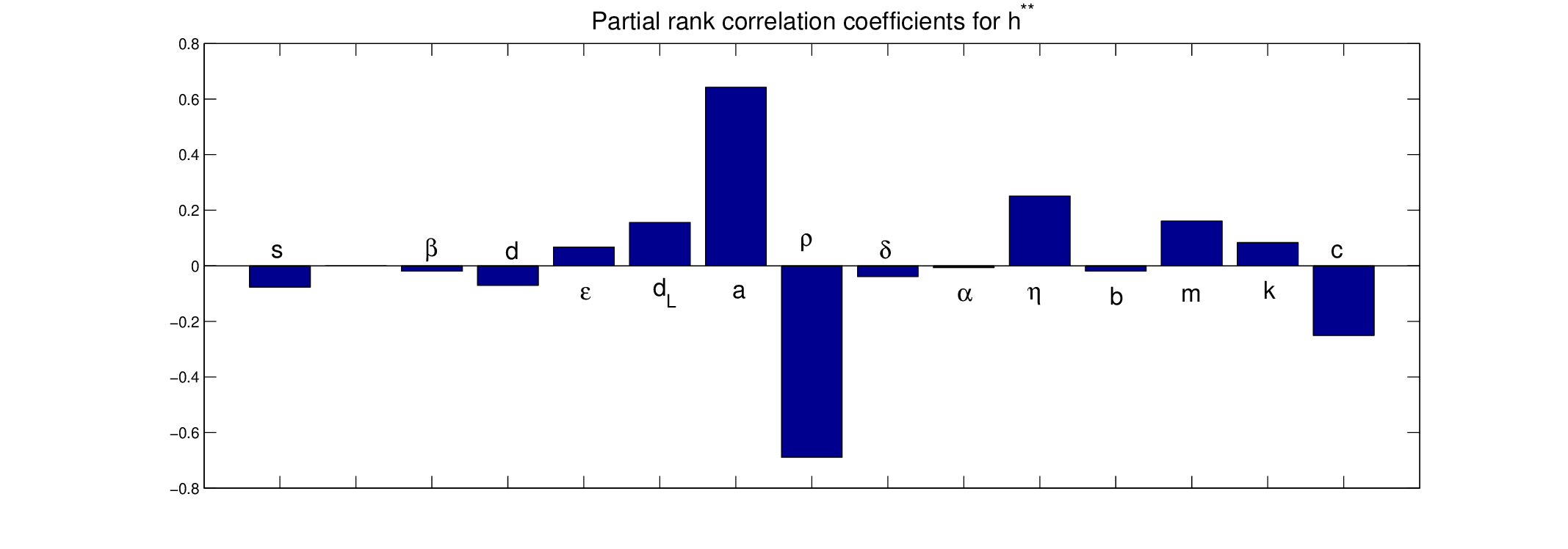}}}
 \caption{
\footnotesize  Partial rank correlation coefficients illustrating
the dependence of $h^{**}$ on each parameter.  
 }\label{F51}
\end{center}
 \end{figure}

\begin{figure}[!h]
\begin{center}
{\rotatebox{0}{\includegraphics[width=0.96 \textwidth,
height=96mm]{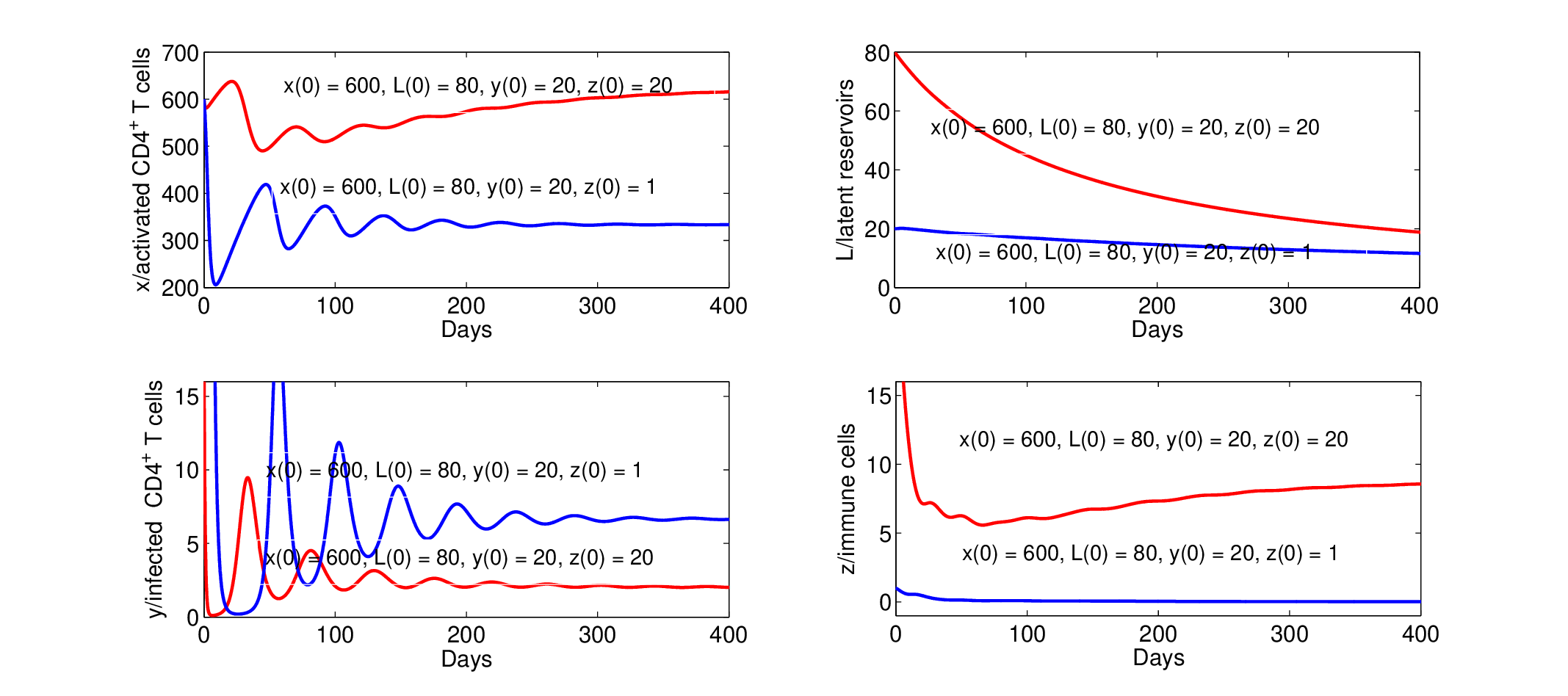}}}
 \caption{
\footnotesize   The bistability of system (\ref{e1}). Here,
$L_{max}=50.$ The initial values are $x(0)=600$, $L(0)=20$,
$y(0)=20$, $z(0)=1$ (blue) and $x(0)=600$, $L(0)=80$, $y(0)=20$,
$z(0)=20$ (red). The parameter values are listed in Table 5.
}\label{F51}
\end{center}
 \end{figure}

% etc

\end{document}